\documentclass[aps,notitlepage,a4paper,superscriptaddress,11pt]{article}

\usepackage[utf8]{inputenc}
\usepackage[dvips]{color}
\usepackage[toc]{appendix}
\usepackage[hang,flushmargin]{footmisc} 
\usepackage{physics}
\usepackage{titlesec,titletoc,ntheorem-hyper,authblk,abstract,latexsym,graphicx,microtype,booktabs,cleveref,amsmath,fancyhdr,wrapfig,array,amssymb,geometry,appendix,cite}

\newcommand{\be}{\begin{equation}}
\newcommand{\ee}{\end{equation}}
\newcommand{\bea}{\begin{eqnarray}}
\newcommand{\eea}{\end{eqnarray}}
\newcommand\blfootnote[1]{%
  \begingroup
  \renewcommand\thefootnote{}\footnote{#1}%
  \addtocounter{footnote}{-1}%
  \endgroup
}

\title{\textbf{Evolution operator for time-dependent non-Hermitian Hamiltonians}}

\author {\textbf{Bijan Bagchi}}

\affil {\it{Department of Physics, Shiv Nadar University, Gautam Buddha Nagar, Uttar Pradesh 201314, India}}

\date{\vspace{-6ex}}

\begin{document}
	\maketitle	
	
\begin{abstract}

The evolution operator $U(t)$ for a time-independent parity-time-symmetric systems is well studied in the literature. However, for the non-Hermitian time-dependent systems, a closed form expression for the evolution operator is not available. In this paper, we make use of a procedure, originally developed by A.R.P. Rau [Phys.Rev.Lett, 81, 4785-4789 (1998)], in the context of deriving the solution of Liuville-Bloch equations in the product form of exponential operators when time-dependent external fields are present, for the evaluation of $U(t)$ in the interaction picture wherein the corresponding Hamiltonian is time-dependent and in general non-Hermitian. This amounts to a transformation of the whole scheme in terms of addressing a nonlinear Riccati equation the existence of whose solutions depends on the fulfillment of a certain accompanying integrabilty condition.

\blfootnote{
    E-mail: {bbagchi123@gmail.com}}
\end{abstract}

\section{Introduction}

Parity-time (PT) symmetric quantum mechanical Hamiltonians, of which the time-reversal is an anti-linear operator, form a distinct sub-class of a wider branch of non-Hermitian Hamiltonians \cite{Ben1, Mos1}. Such Hamiltonians are of interest because a system possessing an exact PT-symmetry generally preserves the reality of their bound-state eigenvalues \cite{Ben2}. Should PT be broken, the eigenvectors cease to be simultaneous eigenfunctions of the joint PT-operator, and as a result, complex eigenvalues spontaneously turn up in conjugate pairs. We refer to such a system as belonging to a PT-broken phase. The PT- transition causes a system to cross over from an equilibrium to a non-equilibrium state. 
During the past few years, the relevance of PT-structure has been noticed
in various optical systems wherein balancing gain and
loss is an interesting issue toward experimental realization of
PT-symmetric Hamiltonians \cite{Mus, Makr, Guo}.

A $PT$-symmetric system is often thought to evolve in a manner wherein the accompanying time evolution of the state vector
is unitary with respect to the $CPT$ inner product.
However, up until now there has been no evidence of experimental support of any physical system for which the time evolution proceeds through a $CPT$-inner product. Of course,
for a non-Hermitian, PT-symmetric Hamiltonian, when looked upon as an open quantum system, loss of unitarity is not a big issue irrespective of whether the spectrum of H is 
purely real or supports complex conjugate pairs of eigenvalues. Recall that in conventional quantum mechanics the concept of Hermiticity holds with respect to the Hamiltonian of a closed system resulting in the reality of the energy spectrum (see, for instance, \cite{Jog1}) .

In the literature there have been efforts to study
several classes of non-Hermitian Hamiltonians even for time-dependent situations \cite{Dut, Yuce, Fring1, Mos2, Gong, Dey, Mam, Fring2, Ramos, Fring3} including the time periodic cases in the Rabi problem \cite{Jog2, Jog3}.
It was recently pointed out \cite{Fring2} that introducing explicit time-dependence into the coupling parameters can render the corresponding Hamiltonian to be physically viable and gave justifications toward the existence of such a possibility in a quantum mechanical context. This motivates us
to have a fresh look at the class of such Hamiltonians, even if non-Hermitian, for which an explicit form of the evolution operator can be constructed in the set-up of an interaction picture. Here, as we shall see, the adoption of the technique implemented by Rau several years ago \cite{Rau1}, in the context of solving Liouville-Bloch equations, proves to be extremely facilitating (see also \cite{Rau2}). We note in passing that in several quantum mechanical pictures an equivalent Hermitian counterpart of the underlying non-Hermitian Hamiltonian exists through the use of the so-called Dyson map that transforms a non-Hermitian 
Hamiltonian into an equivalent Hermitian form \cite{Sch, Kre, Bqr}.

Our paper is organized as follows. In section 2, we run through some of the basic equations of the time-dependent quantum mechanics formalism including writing down the standard expression of the evolutionary operator. In section 3, we summarize the basic features of a two-level PT-symmetric system and express the associated matrix Hamiltonian in terms of the Pauli matrices for use in the later sections. In section 4, we take up the derivation of the evolutionary operator $U(t)$ of such a system through the use of Baker-Campbell-Hausdorff (BCH) formula and note that the complete solvability of the problem requires coming to terms with a nonlinear Riccati equation. In section 5, we address the evaluation of $U(t)$ for a time-dependent two-level spin system adopting a similar strategy as in the previous section. Finally, in Section 6, we provide a brief summary of our work.

\section{Time-dependent quantum mechanics}

Consider a time-dependent non-Hermitian Hamiltonian $H(t)$ having a Hermitian counterpart $h(t)$. The corresponding Schrodinger equations that these Hamiltonians obey are 

\begin{equation}
H(t)\psi(t) = i\hbar\partial_t \psi(t), \quad h(t)\phi(t) = i\hbar \partial_t \phi(t)
\end{equation}
Dyson's map connects $\psi (t)$ to $\phi (t)$ through

\begin{equation}
\phi(t)=\eta(t)\psi(t)
\end{equation}
and provides the following link between $H(t)$ and $h(t)$ in the manner

\begin{equation}
h(t)=\eta(t)H(t)\eta^{-1} (t) + i\hbar \partial_t \eta(t)\eta^{-1} (t)
\end{equation}

For a general quantum mechanical Hamiltonian $\tilde{H}$  \cite{Sakurai}, the Schrodinger equation of the time evolution operator $U(t,t_0)$ is given by 

\begin{equation}
i\hbar\dot{U}(t,t_0) =\tilde{H} U(t,t_0), \quad U(t_0, t_0)=I, \quad t>t_0
\end{equation}
where if the Hamiltonian does not depend on time a solution of the above equation emerges as 

\begin{equation}
U(t,t_0)= exp[\frac{-i\tilde{H}(t-t_0)}{\hbar}]
\end{equation}
The Hamiltonian is considered to be Hermitian because in a standard quantum mechanical formalism the observables correspond to the expectation value of Hermitian operators. 


However, if the Hamiltonian depends on time, as is the case of the interaction picture, the solution is given by

\begin{equation}
U(t,t_0)= T(exp[\frac{-i}{\hbar} \int_{t_0}^t \tilde{H}(t') dt'])
\end{equation}
where $T$ stands for the time-ordering operator and the Hamiltonian $\tilde{H}$ is time-dependent: $\tilde{H}=\tilde{H}(t)$ and is considered Hermitian. 
The standard approach to tackle $U(t)$ is to consider its iterative evaluation by solving (6) comprehensively in terms of nested time-integrations over different categories of time-ordered product. In the following we will set $\hbar =1$. 

In this context, it is useful to note that in attempting to solve the evolution equation, the Lie algebraic approach, based on the Wei-Norman theorem has also been effective \cite{Wei}. In order to reduce the evolution operator in a disentangled form here too one needs to solve a Riccati equation \cite{Kenmore, Guerrero, Cruz}. 

Although for the determination of the evolution operator $U$ in a non-Hermitian scenario 
the explicit unitariness is violated, we can effectively use the technique developed in \cite{Rau1, Rau2} 
by postulating for $U(t)$ a representation in terms of the product of a finite number of exponential operators which close up on the use of BCH expansion for certain classes of commutators. The closure ensures that on moving the relevant operators judiciously, the exponentials stand to the extreme right and facilitate comparison with the given non-Hermitian Hamiltonian a rather straightforward task. Actually, such a procedure allows us to obtain a set of consistency conditions which, under certain advantageous situations, can be completely solved in terms of a nonlinear Riccati equation.

\section{A two-dimensional PT-symmetric system}

Let us concentrate on the following non-Hermitian but a PT-symmetric two-level system \cite{Ben3} 
 
\begin{equation}
\hat{H}=\begin{pmatrix}
re^{i\theta} & s\\
s & re^{-i\theta} 
\end{pmatrix}, \quad \hat{H} \neq \hat{H}^\dagger
\end{equation}
where ${H}^\dagger$ is the Hermitian conjugate of $\hat{H}$ and the three parameters $r,s,\theta$ are real.\\

The eigenvalues of $\hat{H}$ are $\lambda_{\pm} = r\cos\theta \pm\sqrt{s^2-r^2\sin^2\theta}$ which stay real when the inequality $s^2>r^2\sin^2\theta$ holds. The above Hamiltonian commutes with the $PT$ operation i.e. $[\hat{H},PT]=0$ with 

\begin{equation}
   P=
  \left[ {\begin{array}{cc}
   0 & 1 \\
   1 & 0 \\
  \end{array} } \right]
\end{equation}
and $T$ pointing to the usual complex conjugation operation.\\

The simultaneous eigenstates of $H$ and $PT$ are

\begin{equation}
\ket{\psi_{+}} = \frac{1}{\sqrt{2\cos\alpha}}
\begin{pmatrix}e^{\frac{i\alpha}{2} }\\ e^{-\frac{i\alpha}{2} }\end{pmatrix}
\end{equation}
and

\begin{equation}
\ket{\psi_{-}} = \frac{1}{\sqrt{2\cos\alpha}}
\begin{pmatrix}e^{-\frac{i\alpha}{2} }\\ -e^{\frac{i\alpha}{2}} \end{pmatrix}
\end{equation}
where $\sin\alpha = \frac{r}{s} \sin\theta$.

It is important to realize that if the above condition for real eigenvalues is met then $PT$ is unbroken. If contrary is the case, then the states $\ket{\psi_{+}}$ and $\ket{\psi_{-}}$ are no longer eigenstates of $PT$ because $\alpha$ becomes imaginary and points to PT entering a zone of broken phase. Indeed the $PT$ operations are $PT\ket{\psi_{+}}=\ket{\psi_{-}}$ and vice versa.

From now onward we are going to work with the following form for $H$

\begin{equation}
\hat{H}=r\cos\theta I +ir\sin\theta \sigma_z +s\sigma_x
\end{equation}
where $I$ is the identity matrix and $\sigma_{\pm} = \sigma_x \pm i\sigma_y$, the Pauli matrices $\sigma_x, \sigma_y$ along with $\sigma_z$ being

\begin{equation}
\sigma_x=\begin{pmatrix}
0 & 1\\
1 & 0 
\end{pmatrix}, \quad \quad \sigma_y=\begin{pmatrix}
0 & -i\\
i & 0 
\end{pmatrix}, \quad \quad \sigma_z=\begin{pmatrix}
1 & 0\\
0 & -1 
\end{pmatrix}
\end{equation}\\

The PT-symmetric evolution operator for the Hamiltonian $(11)$ has been evaluated in a closed form \cite{Ben4, Assis} by the use of the formula $(5)$ in connection with the solvability of the Brachistochrone problem \cite{Ben4}.

For the time dependent version of $\hat{H}$ we adjust the coefficient parameters to re-cast it in the form 

\begin{equation}
\hat{H}= \nu I +i\kappa (t) \sigma_z +\lambda (t)\sigma_x \equiv \nu I +i\kappa (t) \sigma_z +\frac{\lambda (t)}{2}(\sigma_+ +\sigma_-)
\end{equation}
where $\nu$ is assumed to be a real constant and $\kappa$, $\lambda$ are taken to be, in general, real and continuous functions of $t$. If the latter are arbitrary then $\hat{H}$ is of course non-Hermitian but for situations when $\kappa (t)$ and $\lambda (t)$ are invariant under $t->-t$, PT-symmetry will still apply.
In the following, our task would be to determine the evolution operator $U(t)$ associated with $\hat{H}$.

\section{Evolution operator of the time-dependent non-Hermitian Hamiltonian $\hat{H(t)}$}

Let us project $U$ in the form 

\begin{equation}
U(t)=e^{-ia(t)}e^{ib(t)\sigma_+}e^{ic(t)\sigma_-}e^{d(t)\sigma_z}
\end{equation}
where the real functions $a(t), b(t), c(t)$ and $d(t)$ are to be determined, analogous to the arrangement of the exponentials in \cite{Rau1}. However, it is to be remarked that $U(t)$ is non-unitary here because of the choice of the last factor $d(t)$ in the right side. Taking the time derivative gives us the expression

\begin{equation}
i\dot{U}(t)= \dot{a}IU -\dot{b}\sigma_{+} U -(\dot{c}+2c\dot{d})e^{-ia}e^{ib\sigma_+} \sigma_{-} e^{ic\sigma_-}e^{d\sigma_z} +i \dot{d}e^{-ia}e^{ib\sigma_+} \sigma_z e^{ic\sigma_-}e^{d\sigma_z}
\end{equation}

A bit of manipulations on the last two terms of the right side through the use of the BCH formula namely that for two operators $A$ and $B$ the expansion

\begin{equation}
e^A Be^{-A}=B +[A,B]+\frac{1}{2!}[A,[A,B]]+...
\end{equation}
implies

\begin{equation}
e^A B=(B +[A,B]+\frac{1}{2!}[A,[A,B]]+...)e^A
\end{equation}
we have the relationships

\begin{equation}
i(\dot{c}+2c\dot{d})e^{-ia}e^{ib\sigma_+} \sigma_{-} e^{ic\sigma_-}e^{d\sigma_z} = i(\dot{c}+2c\dot{d})(\sigma_- +4ib\sigma_z +4b^2\sigma_+)U
\end{equation}
and 

\begin{equation}
\dot{d}e^{-ia}e^{ib\sigma_+} \sigma_z e^{ic\sigma_-}e^{d\sigma_z}=\dot{d} (\sigma_z -2ib\sigma_+)U
\end{equation}
In the above we used the commutation rules 

\begin{equation}
[\sigma_z,\sigma_{\pm}]=\pm2\sigma_{\pm}, \quad [\sigma_+,\sigma_-]=4\sigma_z
\end{equation}
Thus (15) is converted to the form 

\begin{equation}
i\dot{U}(t)= [\dot{a}I +i(-4b\dot{c} +\dot{d} -8bc\dot{d})\sigma_z -(\dot{b} +4b^2\dot{c} -2b\dot{d} +8b^2c\dot{d})\sigma_+ -(\dot{c}+2c\dot{d})\sigma_-]U
\end{equation}

Identifying the items inside the squared-brackets in the right-side with the time-dependent Hamiltonian $\hat{H}$ and comparing it with $(13)$ gives us a set of four equations by seeking consistencies between them namely,

\begin{equation}
\dot{a} =\nu
\end{equation}
\begin{equation}
-4b\dot{c} +\dot{d} -8bc\dot{d} =\kappa(t)
\end{equation}
\begin{equation}
\dot{b}+4b^2\dot{c} -2b\dot{d} +8b^2c\dot{d} =-\frac{\lambda(t)}{2}
\end{equation}
\begin{equation}
\dot{c} +2c\dot{d} = -\frac{\lambda(t)}{2}
\end{equation}

With $\nu$ being devoid of any time-dependence, the parameter $a$ can be determined by direct integration which turns out to be a linear function of $t$.  The equation for $b$ can be found out by first combining (24) and (25) to get 

\begin{equation}
\dot{b} -2\lambda b^2 -2b\dot{d} =-\frac{\lambda}{2}
\end{equation}
and then through (23) and (25) arriving at

\begin{equation}
\dot{d} +2b\lambda  =\kappa
\end{equation}
From the last two equations (26) and (27) we can eliminate  $\dot{d}$ to get a first-order ordinary differential equation

\begin{equation}
\dot{b}-2\kappa b +2\lambda b^2 = -\frac{\lambda}{2}
\end{equation}
The above equation can be recognized to be in the nonlinear Riccati form.

To tackle such an equation it is interesting to consider the underlying integrability condition as recently pointed out by Mak and Harko \cite{Mak}. They observed that the general solution of an equation of the form (28) depends upon the coefficients of the equation satisfying an auxiliary condition. In the present context the latter reads in terms of a suitably defined generating function $f(t)$

\begin{equation}
\lambda +\frac{d}{dt}(\frac{-2\kappa +\sqrt{f}}{2\lambda})+\frac{4\kappa^2 -f}{4\lambda}=0
\end{equation}
and a similar one corresponding to the negative sign of the square-root. While for arbitrary $\kappa (t)$ and $\lambda (t)$ this equation appears to be rather complex to allow the Riccati equation to be solved in a closed form, certain special cases could work. For instance, when the derivative term in $(29)$ is disregarded by having

\begin{equation}
\frac{-2\kappa +\sqrt{f}}{2\lambda} = \mbox{constant}
\end{equation}
one can derive plausible relations between $\kappa$ and $\lambda$ for which (29) holds. To pursue this point a little further let us take the constant to be $c (\neq 1)$. Then from $(30)$ $f$ turns out to be 

\begin{equation}
f=4(\kappa +c\lambda)^2
\end{equation}
while $(29)$ implies 

\begin{equation}
f=4(\lambda^2 +\kappa^2)
\end{equation}
Equating the above two expressions for $f$ we easily find that $\lambda$ is related to $\kappa$ as 

\begin{equation}
\lambda = \frac{2c\kappa}{1-c^2}, \quad c\neq 1
\end{equation}
implying

\begin{equation}
f=4\kappa^2(\frac{1+c^2}{1-c^2})^2
\end{equation}
Once $f$, $\lambda$ and $\kappa$ are known, the general solution of $b$ satisfying the Riccati equation $(28)$ is known too as given by a rational expression. The form is too complicated to be reproduced here and we refer to \cite {Mak} for a further study (see also \cite{Rosu}).

Once $b$ is known,  a solution of $d$ is supposed to follow from (27) and as a consequence the parameter $c$ is can be found out from either (25) or (23). In this way, the evolution operator $U$ for the non-Hermitian Hamiltonian $(13)$ is determined completely.

\section{Evolution operator of the  time-dependent two-level spin model}

The Hamiltonian for a two-level spin model is modeled by

\begin{equation}
\hat{H}= -\frac{1}{2}[\omega I +\lambda \sigma_z +i\kappa\sigma_x]
\end{equation}
where $\omega, \lambda, \kappa $ are real coupling parameters. By diagonalization the reality of the energy spectrum corresponds to the inequality $|\lambda|>|\kappa|$. The anti-linear symmetry operator corresponds to $PT$.

Fring and Frith \cite{Fring2} chose the time-dependent counterpart of $\hat{H}$ to be

\begin{equation}
\hat{H}= -\frac{1}{2}[\omega I +\alpha\kappa(t) \sigma_z +i\kappa(t)\sigma_x] \equiv -\frac{1}{2}\omega I -\frac{\alpha}{2}\kappa(t) \sigma_z -i\frac{\kappa(t)}{4}(\sigma_+ +\sigma_-)
\end{equation}
where $\alpha$ is real. Notice that the main difference of the above form of the Hamiltonian with the one in $(13)$ is that, apart from an overall sign factor, the presence of the imaginary number $i$ in the coefficients of $\sigma_z$ and $\sigma_x$ is exchanged. In the simplified model studied in \cite{Fring2} the coefficient of $\sigma_z$ has been set equal to $\alpha\kappa(t)$. Subsequently, it was shown that the time-dependent coefficient $\lambda$ is related to the solution of the Ermakov-Pinney equation.

To work out the time evolution operator associated with the Hamiltonian $(36)$ we need to adopt a slightly different form for $U(t)$ because of the position of the imaginary number $i$ being different from $(13)$. To this end, let us take a non-unitary $U$ as given by the following succession of exponentials

\begin{equation}
U(t)=ie^{a(t)}e^{ib(t)\sigma_+}e^{ic(t)\sigma_-}e^{d(t)\sigma_z}
\end{equation}
where the real functions $a(t), b(t), c(t)$ and $d(t)$ are to be determined in the spirit of what we did in the previous section.

Making use of BCZ formula to move the Pauli matrices in the right order gives

\begin{equation}
i\dot{U}(t)= [i\dot{a}I +i(-4b\dot{c} +\dot{d} -8bc\dot{d})\sigma_z -(\dot{b} +4b^2\dot{c} -2b\dot{d} +8b^2c\dot{d})\sigma_+ -(\dot{c}+2c\dot{d})\sigma_-]U
\end{equation}

Comparing with the form $(4)$ to identify the Hamiltonian as the one given by $(36)$, where we adjust a factor of $i$, 
gives us the following set of four equations 

\begin{equation}
\dot{a} =-\frac{1}{2} \omega
\end{equation}
\begin{equation}
-4b\dot{c} +\dot{d} -8bc\dot{d} =-\alpha\frac{\kappa(t)}{2}
\end{equation}
\begin{equation}
\dot{b}+4b^2\dot{c} -2b\dot{d} +8b^2c\dot{d} =-\frac{\kappa(t)}{4}
\end{equation}
\begin{equation}
\dot{c} +2c\dot{d} = -\frac{\kappa(t)}{4}
\end{equation}

These equations readily furnish the counterpart equation to $(28)$ namely

\begin{equation}
\dot{b} +\alpha\kappa b +\kappa b^2 =  -\frac{\kappa(t)}{4}
\end{equation}
which is again of the Riccati type and can be handled in a similar way as discussed earlier. Knowing $b$ one can determine, in principle, the other two function $c(t)$ and $d(t)$.

\section{Summary}

In this work, we attempted to evaluate the evolution operator for a couple of time-dependent non-Hermitian Hamiltonians. The first one we studied has its roots in the general class of PT -symmetric $2 \times 2$ systems with time-dependence explicitly inserted among its coupling constants. By writing the evolution operator as a product of exponential factors defined in terms of time-dependent coefficient functions, but in a non-unitary form, we showed that these functions could be determined by exploiting the closed operator algebra of the Pauli matrices. Of course, this required a quadratic Riccati equation to be solved which in turn called for the fulfillment of an integrability conditon. Although somewhat complicated in nature we pointed out a simple case when the Riccati equation can indeed be handled straightforwardly. Our second model of inquiry was that of a time-dependent two-level spin model. Here we had to take the evolution operator in a slightly different form. We found that in this case too one of the coefficient functions satisfies a nonlinear equation in the Riccati form whose knowledge sheds light on the other couplings. The form of the Riccati equation turned out to be similar to the previous one.

\newpage

\section*{Acknowledgments}

I thank Prof. A.R.P. Rau for several enlightening correspondences. I also thank Prof. Y.N. Joglekar for a valuable comment and making useful suggestions, and Prof. A. Fring for bringing reference $[17]$ to my attention. I am grateful to Prof. M. Maamache for pointing out an error in the earlier version of the manuscript and to an anonymous referee for making certain constructive advises.

\end{document}